\documentclass[prd,showpacs,preprintnumbers,nofootinbib,aps,twocolumn]{revtex4}
\usepackage{subeqnarray}
\usepackage{epsfig,amsmath,amssymb}
\usepackage{mathrsfs}
\usepackage[usenames,dvipsnames]{color}
\usepackage[pagebackref=true, colorlinks=true]{hyperref}
\definecolor{redish}{rgb}{0.7,0.2,0.0}  % color defined in (r=red,g=green,b=blue) model
\definecolor{bluish}{rgb}{0.2,0.5,0.8}
\hypersetup{linkcolor=redish,          % color of internal links
                  citecolor=blue,        % color of links to bibliography
                  filecolor=magenta,      % color of file links
                  urlcolor=bluish}          % color of the url
\DeclareFontFamily{U}{rsfs}{}         % Formal Script            %
\DeclareFontShape{U}{rsfs}{m}{n}{<5> rsfs5 <6><7> rsfs7          %
  <8><9><10><10.95><12><14.4><17.28><20.74><24.88> rsfs10}{}     %
\DeclareMathAlphabet{\mathfs}{U}{rsfs}{m}{n}                     %
\newcommand{\mfs}[1]{\mathfs {#1}}                              %

\newcommand{\ba}{\nopagebreak[3]\begin{eqnarray}}
\newcommand{\ea}{\end{eqnarray}}
\newcommand{\bii}{\begin{itemize}}
\newcommand{\eii}{\end{itemize}}
\newcommand{\nn}{\nonumber}

\newcommand{\sO}{{\mfs O}}

\newcommand{\f}{\frac}

\def \g{\gamma}
\def \e{\epsilon}

\def \lm{\lambda}

\def \O{\Omega}

\def \({\left(}
\def \){\right)}
\def \[{\left[}
\def \]{\right]}
\begin{document}
\title{Non-extensive Statistical Mechanics and Black Hole Entropy From Quantum Geometry}
\author{Abhishek Majhi}%
\email{abhishek.majhi@gmail.com}
\affiliation{Instituto de Ciencias Nucleares\\
Universidad Nacional Autonoma de Mexico\\
A. Postal 70-543, Mexico D.F. 04510, Mexico}

\begin{abstract}
Using non-extensive statistical mechanics, the Bekenstein-Hawking area law is obtained from microstates of black holes in loop quantum gravity, for arbitrary real positive  values of the Barbero-Immirzi parameter$(\g)$. The arbitrariness of $\g$ is encoded in the strength of the ``bias'' created in the horizon microstates through the coupling with the quantum geometric fields exterior to the horizon. An experimental determination of $\g$ will fix this coupling, leaving out the macroscopic area of the black hole to be the only free quantity of the theory. 
\end{abstract}
\maketitle

\section{Introduction}
Loop quantum gravity(LQG) provides us with an estimate of the microstates of a black hole with given classical area$(A)$, albeit at the kinematical level, which leads to a precise computation of its entropy\cite{qg2,km98}. The whole procedure is completed in two steps. Step 1: One deals with the field dynamics on the horizon to unravel the nature of the Hilbert space, hence the microstates, associated with the same. Step 2: Having the estimate of the microstates at hand, one applies the statistical mechanics to calculate the entropy$(S)$.

 The calculation yields $S=(\lm_0/\g) A$, where $\g$ is known as the Barbero-Immirzi parameter\cite{bar,imm1,imm2}, $\lm_0$ is a numerical constant resulting from the underlying statistical mechanics. Then, demanding that $S$ be given by the Bekenstein-Hawking area law(BHAL) i.e. $A/4$\cite{bek,haw}, the value of $\g$ is suitably fixed. Now, the parameter $\g$ represents a one-parameter family of canonical transformation of the canonical variables of the classical theory i.e. for every value of $\g$ the classical equations of motion of general relativity are valid. And, in the quantum theory $\g$  represents a quantization ambiguity of the theory like the $\theta$-parameter of quantum chromodynamics. For every real and positive\footnote{$\g$ should be real and positive in the quantum theory because it appears as a multiplicative constant in the area spectrum of the black hole\cite{qg2}.} value of $\g$ there is a valid quantum theory, but they are unitarily inequivalent. In principle, if the derivation of the black hole entropy is correct, then the BHAL should follow for all real and positive values of $\g$. Hence, the necessity of choosing $\g$ by hand just for the sake of obtaining the BHAL is considered as a drawback of this approach of black hole entropy calculation\cite{jac}. It is suggested that one should obtain the $S=A/4$ for arbitrary $\g$. Hence, a physical explanation behind the tuning of $\g$ and obtaining the BHAL for arbitrary $\g$, both will be value additions to the concerned literature. We achieve these goals by introducing a statistical mechanics or rather definition of entropy other than the one, namely Boltzmann entropy, used in the standard literature to calculate black hole entropy from LQG. So, let us provide the physical motivation behind doing that on the first place.

\section{Motivation}
  The physical motivation behind  changing the standard definition of entropy comes from an observation regarding the two steps involved in the black hole entropy calculation in LQG.% the following fact. The nature of field dynamics on the black hole horizon, especially the coupling between the horizon gauge field theory with the bulk geometry, suggest that  a bias needs to be  incorporated in the microstates  on the horizon while defining the entropy. Such a bias is not incorporated in the standard definition of entropy. So, let us discuss briefly the two steps of the procedure of black hole entropy calculation in LQG, which will bring forth the fact that there is a lack of self-consistency in these two steps of the procedure. This will serve as the motivation to introduce a new definition of entropy in the first place.

Step 1 consists of the exploration of the quantum field dynamics on the horizon.  Classically, it has been shown for the case of Schwarzschild black hole that the field equations on the horizon can be derived from the action of a CS theory coupled to an external source\cite{km11}
\ba 
S_{CS}~&=& ~{\frac k { 4\pi}} \int^{}
\e^{abc}_{} \left( A^{I}_a \partial^{}_b A^{I}_c ~+ ~ {\frac  1 3}
~\e^{IJK}_{}  A^{I}_a A^{J}_b A^{K}_c  \right)  \nn\\
&& ~~~~~~~~~~~~~~~~~~+~ \int
J^{Ia}_{} A^{I}_a ~~ \label{CSAction+} 
\ea
where $A^{I}_a$ is the CS gauge field, $J^{Ia}_{}$ is the external source that is dual to the bulk soldering form both with respect to the internal indices and the spacetime indices; $I,J,K$ represent the internal gauge index and $a,b,c$ represent the spacetime indices.  The coupling with the external source represents the coupling between the horizon and the external bulk. In the quantum theory these sources represent point-like quantum geometric excitations on the horizon\cite{qg2}. The Hilbert space associated with the black hole horizon is that of a Chern-Simons(CS) theory coupled to these point like sources\cite{km98}. It is this field theoretic view-point that originally led to the estimate of the full microstate counting of the horizon from from the dimension of the Hilbert space of CS theory by Kaul and Majumdar\cite{km98}, using the machineries of topological quantum field theory from Witten's work\cite{wit} and Verlinde's formula from conformal field theory\cite{verlinde}. 

%The precise state counting proceeds in the following way. The quantum description of a cross-section of the black hole horizon is given by a topological 2-sphere with point-like quantum geometric excitations carrying SU(2) spin representations ($j$-s) taking values $1/2, 1, 3/2,\cdots, k/2$, $k$ being the level of the CS theory. These quantum geometric excitations give rise to the quantum area of the 2-sphere via the formula\cite{qg2} $A_{qu}=8\pi\g\sum_{l=1}^N\sqrt{j_l(j_l+1)}$, where we have set Planck length to unity. The microstate counting of the horizon is done by taking into account all such possible spin sequences that yield $A_{qu}=A\pm\sO(1)$ for a given classical area $A$ of the black hole. However, the fact that the total SU(2) charge must vanish in a closed universe, is also taken into account\cite{km98,wit}. 

Step 2 begins with the imposition of a statistical mechanics. The black hole entropy calculation in LQG is based on the well-known Shannon entropy formula\footnote{We shall consider the Boltzmann constant to be unity.}
\ba
S=-\sum_{i=1}^{\O}p_i\ln p_i\label{sh}
\ea
where $p_i$ is the probability of the $i$-th microstate and $\O$ is the total number of microstates of the system under consideration. Then considering that all the possible microstates can occur, a priori, with equal probability, one uses $p_i=1/\O$ for all $i$ in eq.(\ref{sh}) to arrive at the formula 
\ba
S=\ln \O.\label{bg}
\ea 
Since the estimate of this $\O$ is now known from the knowledge of the Hilbert space, the rest is just a mathematical procedure that leads to $S=(\lm_0/\g) A$.

Now, one can easily make the observation that in Step 2, the computation of the entropy from the horizon microstates using eq.(\ref{sh}) inherently considers that the microstates are  unbiased. This is possible only if these states {\it were} completely unaffected by any interaction with some external fields. On the other hand, in Step 1, we can see that the microstates of the horizon are described by a quantum CS theory coupled to point-like sources from bulk quantum geometry. 
There is a gravitational coupling or interaction between the horizon and the bulk. Hence, it seems quite logical to introduce some different statistical mechanics or rather an entropy formula more generalized than eq.(\ref{sh}) to take into account the effect of the coupling between the  horizon and the bulk as a bias in the microstates. Now, the question is how should the microstates be biased. The answer has two physical views:
\\
i) From the statistical mechanical viewpoint, the bias should be such that the entropy calculation from LQG leads to the BHAL.
\\
ii) From the field theoretic viewpoint, the bias should be such that it increases  with the strength of the coupling of the horizon microstates with the bulk geometry. 

In this work we show that these two viewpoints complement each other quite naturally if we introduce a generalization of the Shannon entropy(henceforth to be called as $q$-entropy) to incorporate the effect of a bias in the microstates, use it to calculate the black hole entropy and demand it to yield the BHAL. As a consequence, we obtain the BHAL from the black hole microstates in LQG for arbitrary real positive values of $\g$. Nevertheless, at the end $\g$ should have a fixed value that has to be determined by experimental means. Once we are able to do so, the parameter $q$ will become a function of $A$ which is physically well justified because of the following reason. The coupling strength between the horizon and the bulk is dependent on $A$ as $k=A/4\pi\g$. Since $q$ represents the effect of the bias created in the horizon microstates due to this coupling, it should also depend on $A$. %So, now let us prove these assertions in detail.

%Technically, it may look quite trivial to predict the arbitrariness of $\g$ at the expense of the free parameter $q$, if one begins to calculate the entropy with eq.(\ref{tent}). However, what is nontrivial and quite remarkable is that the precise relation between $\g$ and $q$ (for any given $A$) comes out to be such that the bias in the microstates increases with the strength of their coupling with the bulk geometry. This provides us with a physical justification of the tuning of $\g$ to obtain the BHAL i.e. $\g$ couples the bulk geometry to the horizon degrees of freedom in such a way that the entropy of the black hole comes out to be the BHAL. Such a physical explanation  about the tuning of $\g$ has remained hitherto unknown. 

\section{The $q$-entropy}
Originally, the idea behind the introduction of the notion of $q$-entropy was to incorporate, at the statistical mechanical level, the effect of a bias\footnote{The exact nature of bias in a quantum system can only come through the study of its dynamics. $q$-entropy is a way to incorporate that effect at the statistical mechanical level. This is the reason to put the word `bias' within quotes in the abstract.} in the probabilities of the microstates of the underlying quantum mechanical system\cite{ts}(also, see page 43 of \cite{tbook}). The parameter $q$ is called entropic index. In general we have $0<p_i<1$. Hence, $p_i^q>p_i$ for $q<1$ and $p_i^q<p_i$ for $q>1$. This implies $q<1$ relatively enhances the rare events whose probabilities are close to zero and $q>1$ relatively enhances the frequent events whose probabilities are close to unity. Intrigued by this fact, the $q$-entropy was postulated by Tsallis\cite{ts}, which is given by
\ba
S_q=\frac{(1-\sum_{i=1}^{\O}p_i^q)}{(q-1)}, \label{tent}
\ea
The parameter $q$ is real and in the limit $q\to1$ one recovers eq.(\ref{sh}). To mention, the related branch of statistical mechanics is known as non-extensive statistical mechanics(NESM)\footnote{The nomenclature `non-extensive' is slightly misleading(see page 44 of \cite{tbook}). However, we use it here as this branch of statistical mechanics is well known by this name.} due to its salient features\cite{tbook}. For equal probability we have $p_i=1/\O$ for all $i$. In this case eq.(\ref{tent}) reduces to
\ba
S_q=\ln_q \O\label{bgq}
\ea
where $\ln_qx=(1-x^{1-q})/(q-1)$ is called $q$-logarithm.
The $q$-entropy for a spin sequence $(j_1,\cdots,j_N)$ can be calculated by putting $\O(j_1,\cdots,j_N)=\prod_{l=1}^N(2j_l+1)$ in eq.(\ref{bgq}). The expression comes out to be
\ba
S_q(j_1,\cdots,j_N)=\ln_q\prod_{l=1}^N(2j_l+1)=\f{\[\prod_{l=1}^N(2j_l+1)\]^{1-q}-1}{(1-q)}\label{tsgen1}
\ea
If we consider $j_1=\cdots=j_N=s$ (say), then  eq.(\ref{tsgen1}) reduces to the following form:
\ba
S_q^{(s)}&:=&S_q(N \text{number of spin}~s)\nn\\
&=&\f{[(1+2s)^{(1-q)N}-1]}{(1-q)}\label{tsi}
\ea
which was derived in \cite{iden}. This is the equation which we shall implement to calculate black hole entropy.

%Now, we shall find the $q$-entropy of black holes from the microstates and find the conditions that lead to the BHAL. Before doing that, 

\section{Black hole microstates in LQG}
Now, let us briefly discuss the essential structures of the quantum geometry of black holes\cite{qg2,km98}. The quantum geometry of a cross-section of a black hole horizon in LQG is described by a topological two-sphere with defects, usually called punctures, carrying `spin'\footnote{These `spin' quantum numbers are not to be confused with particle spins. For an elaborate discussion see \cite{amindis}.} quantum numbers endowed by the edges of the spin network that represent the bulk quantum geometry\cite{qg2}. Quantum area of the black hole with spin quantum numbers $j_1,\cdots,j_N$ on $N$ punctures is given by $A_{qu}=8\pi\g\sum_{l=1}^N\sqrt{j_l(j_l+1)}$ and the number of microstates is given by
\ba
\O(j_1,\cdots,j_N)=\prod_{l=1}^N(2j_l+1).\label{jh}
\ea
So, we practically have a system of $N$ statistically independent spins with quantum numbers $(j_1,\cdots,j_N)$. The total number of microstates for a black hole with classical area $A$ is counted by taking into account all possible such spin sequences which gives rise to $A_{qu}=A\pm\sO(1)$. This leads to the BHAL from LQG, using standard statistical mechanics, for a fixed value of $\g$\cite{qg2}. However, for a more precise state counting one can see\cite{km98} which leads to logarithmic corrections\cite{km00}. Since our aim is to obtain BHAL from LQG for arbitrary values of $\g$, we shall work with eq.(\ref{jh}) and leave the case of correction terms in the context of NESM for a future study.

\section{Black hole entropy and $\g$}
\begin{figure}[hbt]
\begin{center}
\includegraphics[scale=0.35]{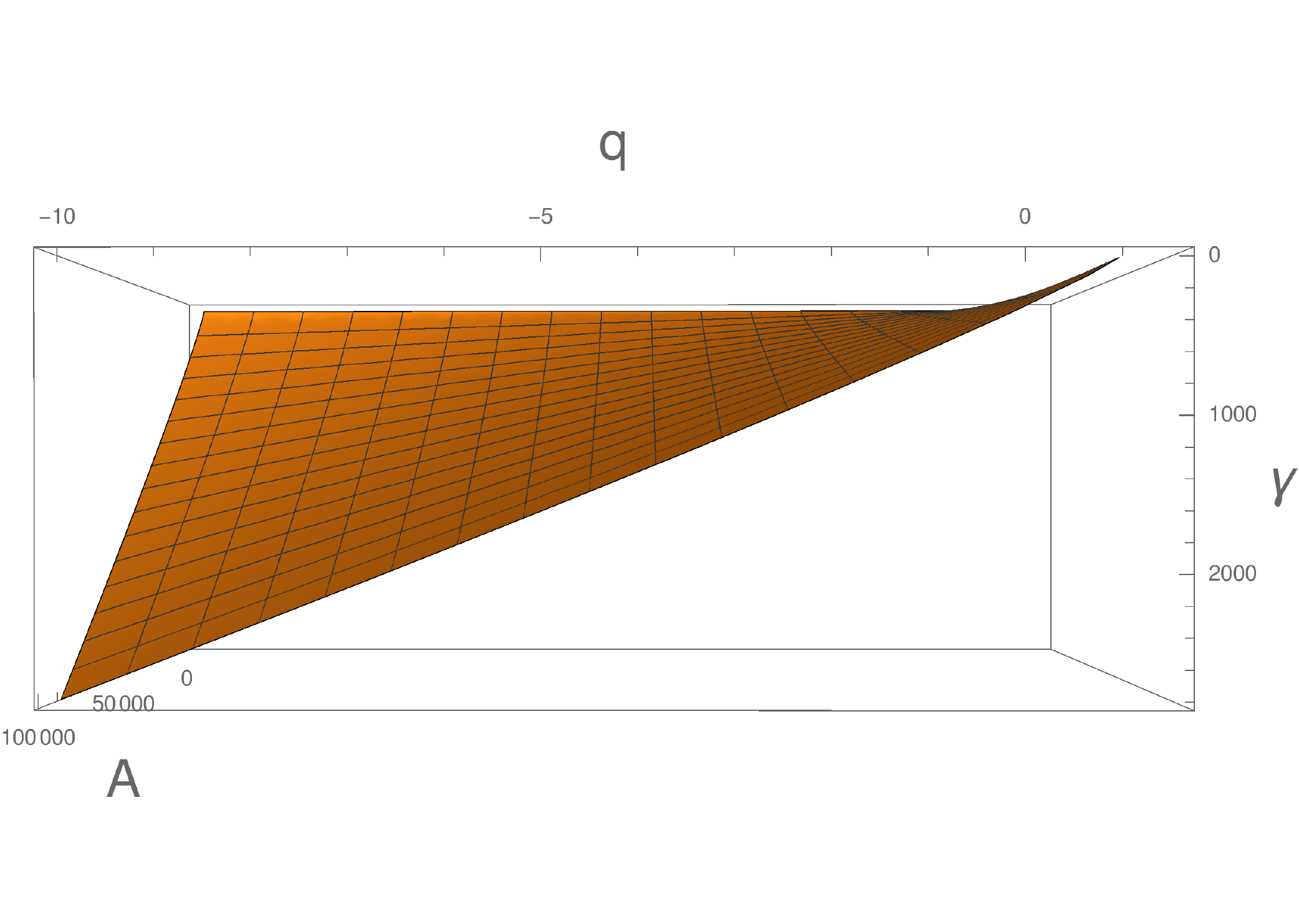}
\caption{\label{tsfig}In this plot we have shown the variation of $\g$ with $A$ and $q$ using eq.(\ref{gq}). The view along $A$-axis is shown to reveal the variation of $\g$ with $q$ explicitly.} 
\end{center}
\end{figure}

%Consequently, the $q$-entropy for the black hole with such a set of punctures is given by eq.(\ref{tsgen}). 
In principle, to find the $q$-entropy for a given $A$ of the black hole we need to take into account all possible such spin sets for which we will have $A_{qu}=A\pm\sO(1)$. Mathematically it means that we have to consider a sum over all possible values of the spins, considering all possible number of punctures $N$, to find the total number of microstates given by $\Omega=\sum_N\sum_{j_1,\cdots,j_N=1/2}^{k/2}\O(j_1,\cdots,j_N)$. Then use that $\O$ in eq.(\ref{bgq}) and extremize the entropy subject to the area constraint. Avoiding this extremization procedure and leaving it for future studies, as a first step of this exercise, we set $j_l=1/2\forall l\in[1,N]$. Physically, this is a good approximation because we are interested in the scenario $N\gg\sO(1)$ which allows us to apply statistical mechanics to the system. Since $1/2\leq j_l\leq k/2~\forall l\in[1,N]$ and we are looking at the quantum states with $A_{qu}=A\pm\sO(1)$ and $A\gg\sO(1)$ for large black holes, $j_1=\cdots=j_N=1/2$ satisfies all the conditions most strongly as it yields $A_{qu}=4\sqrt 3\pi\g N\simeq A$. Therefore, the $q$-entropy of a black hole can be obtained by putting $s=1/2$ in eq.(\ref{tsi}) and it leads to
\ba
S_q^{(1/2)}=\f{[2^{(1-q)N}-1]}{(1-q)}\label{ts1/2}
\ea
Now, for $j_1=\cdots=j_N=1/2$, the area comes out to be $A=4\sqrt 3\pi\g N$. Using this result in eq.(\ref{ts1/2})% we obtain
%\ba
%S_q^{(1/2)}= \f{[2^{(1-q)A/4\sqrt 3\pi\g}-1]}{(1-q)}\label{s1/2}
%\ea
and  demanding that $S_q^{(1/2)}=A/4$, one can solve for $\g$ in terms of $A$ and $q$ to get
\ba
%\f{[2^{(1-q)A/4\sqrt 3\pi\g}-1]}{(1-q)}=\f{A}{4}
\g=\frac{\ln 2 }{\pi\sqrt 3}\cdot\frac{\frac{A}{4}(1-q)}{\ln[1+\frac{A}{4}(1-q)]}\label{gq}
\ea
As a fiducial check, it is easy to see that in the limit $q\to1$, we recover $\g=\ln 2/\pi\sqrt 3$ obtained in \cite{qg2} while calculating black hole entropy  using eq.(\ref{sh}) for $j_1=\cdots=j_N=1/2$. However, the most important issue here is that, for a given value of $A$, we can have arbitrary real positive values of $\g$ depending on $q$. A three dimensional plot of the surface given by eq.(\ref{gq}), viewed along the $A$-axis is shown in fig.(\ref{tsfig}). The cross-section of the surface for $A=10^6$ yields the curve shown in fig.(\ref{A106}), shown as an example, which reveals explicitly the nature of $\g$-$q$ correlation for this particular value of $A$. Similarly, one can take other values of $A$($\gg\sO(1)$) and check the nature of $\g$-$q$ curves to ensure that the overall nature of the curve remains the same. This brings us at a position where we can now quantitatively justify our assertions that we made in the beginning in view of introducing NESM to calculate the BHAL from LQG.
%\vspace{0.1cm}

\section{The bias and the coupling}
 From the statistical mechanical viewpoint, as we depart from $q=1$, the underlying microstates of the associated quantum system become biased. Generically, $p_i$ being the probability of the $i$-th microstate, we have $0<p_i<1$ for any microstate. Hence, $q>1\implies p_i^q<p_i$ which relatively enhances the frequent events i.e. increases the possibility of occurrence of the microstates whose probabilities are closer to unity and $q<1\implies p_i^q>p_i$ which relatively enhances the rare events i.e. increases the possibility of occurrence of the microstates whose probabilities are closer to zero.

%\section{The coupling}
 From the field theoretic viewpoint $1/k=4\pi\g/A$ is the coupling constant which appears in front of the source fields of the bulk geometry that couple to the CS theory on the horizon\cite{km11,qg2}. The field equations on the black hole horizon is given by
\ba
 F^I_{ab}=\frac{1}{k}\cdot\Sigma^I_{ab}=\frac{4\pi\g}{A}\cdot\Sigma^I_{ab}\label{cs}
\ea
where $F^I_{ab}$ is the curvature of the CS gauge fields on the horizon and $\Sigma^I_{ab}$ stands for the pull-back of the soldering form constructed from the bulk tetrads.  It may be noted that the definition of $k$ may differ by some numerical constants in certain references depending on the redefinition of the fields. Eq.(\ref{cs}) is the equation of a CS theory coupled to an external source with the coupling strength being controlled by $4\pi\g/A$. Hence, it is quite explicit that, for a black hole with a given area, $\g$ controls the strength with which  the horizon field theory is sourced and affected by the bulk fields. Quantization of this CS theory coupled to the bulk source fields leads to the Hilbert space of the horizon that provides the estimate of the microstates.% given in eq.(\ref{main}). 

\begin{figure}[hbt]
\begin{center}
\includegraphics[scale=0.4]{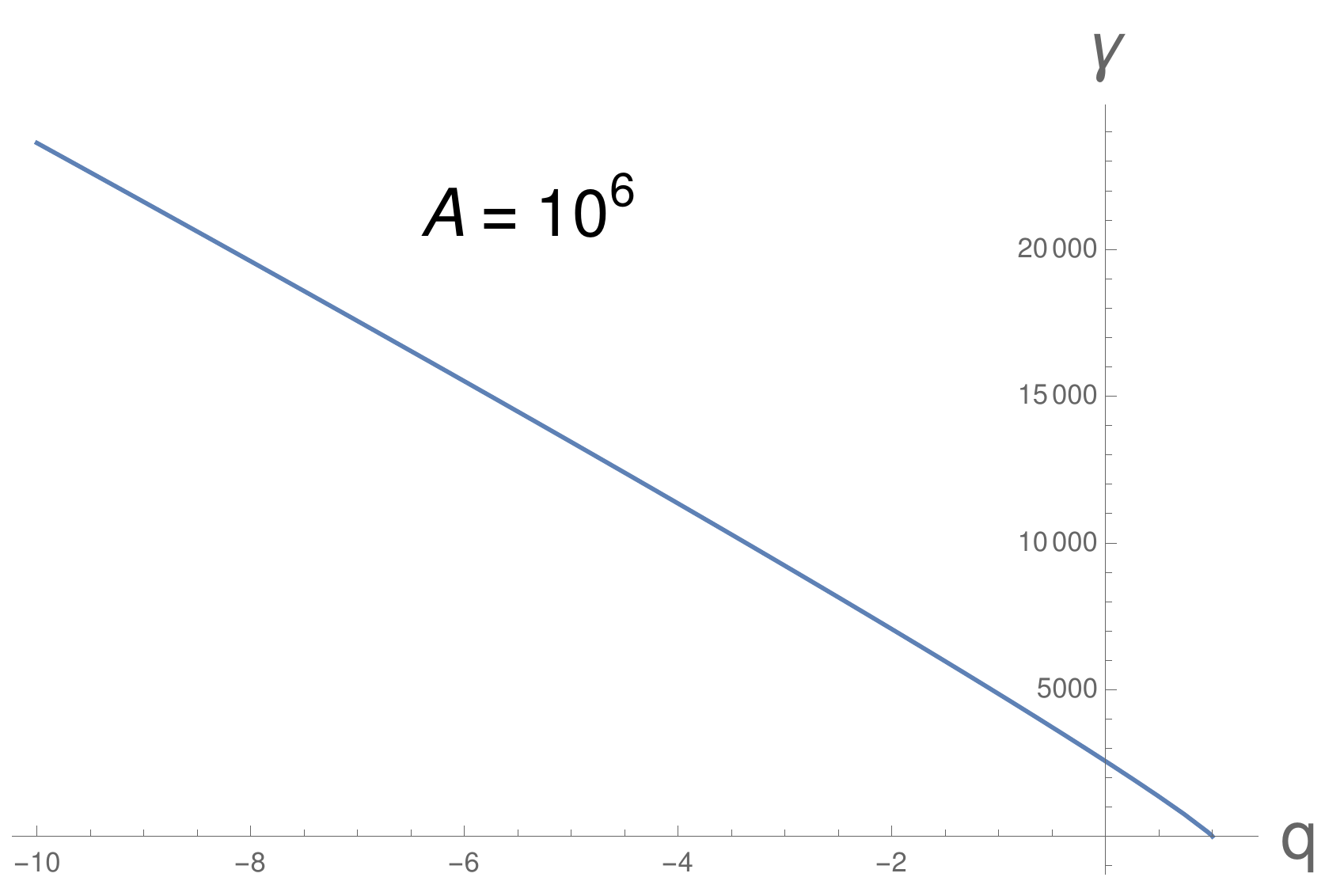}
\caption{\label{A106}In this plot we have shown the variation of $\g$ with $q$ for $A=10^6$ as one example. It shows that $\g$ can take arbitrary positive values depending on $q$ and it increases monotonically with decreasing $q$.  This overall behaviour  of the $\g$ vs $q$ remains the same for arbitrary values of $A\gg\sO(1)$.} 
\end{center}
\end{figure}

%\section{The bias and the coupling}
Thus, for the above viewpoints to complement each other, the microstates need to get more biased (i.e. $q$ departs more from unity) as the coupling between the horizon and the bulk  becomes stronger (i.e. $\g$ increases in magnitude). Quite remarkably, for $q\leq 1$, this is what we can see from fig.(\ref{A106}) which we have obtained by claiming that the entropy computed from LQG be given by the BHAL for $A=10^6$.
The plot shows that the value of $\g$ increases monotonically as $q$ decreases from unity. Since, for a given $A$, $\g\to\infty$ is the strong coupling limit, it implies that the bulk affects  the horizon and biases microstates more strongly as we depart more from $q=1$. Hence, we can conclude that the horizon microstates get biased due to the interaction with the bulk in a precise fashion so as to yield the BHAL. All these provide a physical interpretation of the tuning of $\g$ to obtain the BHAL from LQG.

\begin{figure}[hbt]
\begin{center}
\includegraphics[scale=0.3]{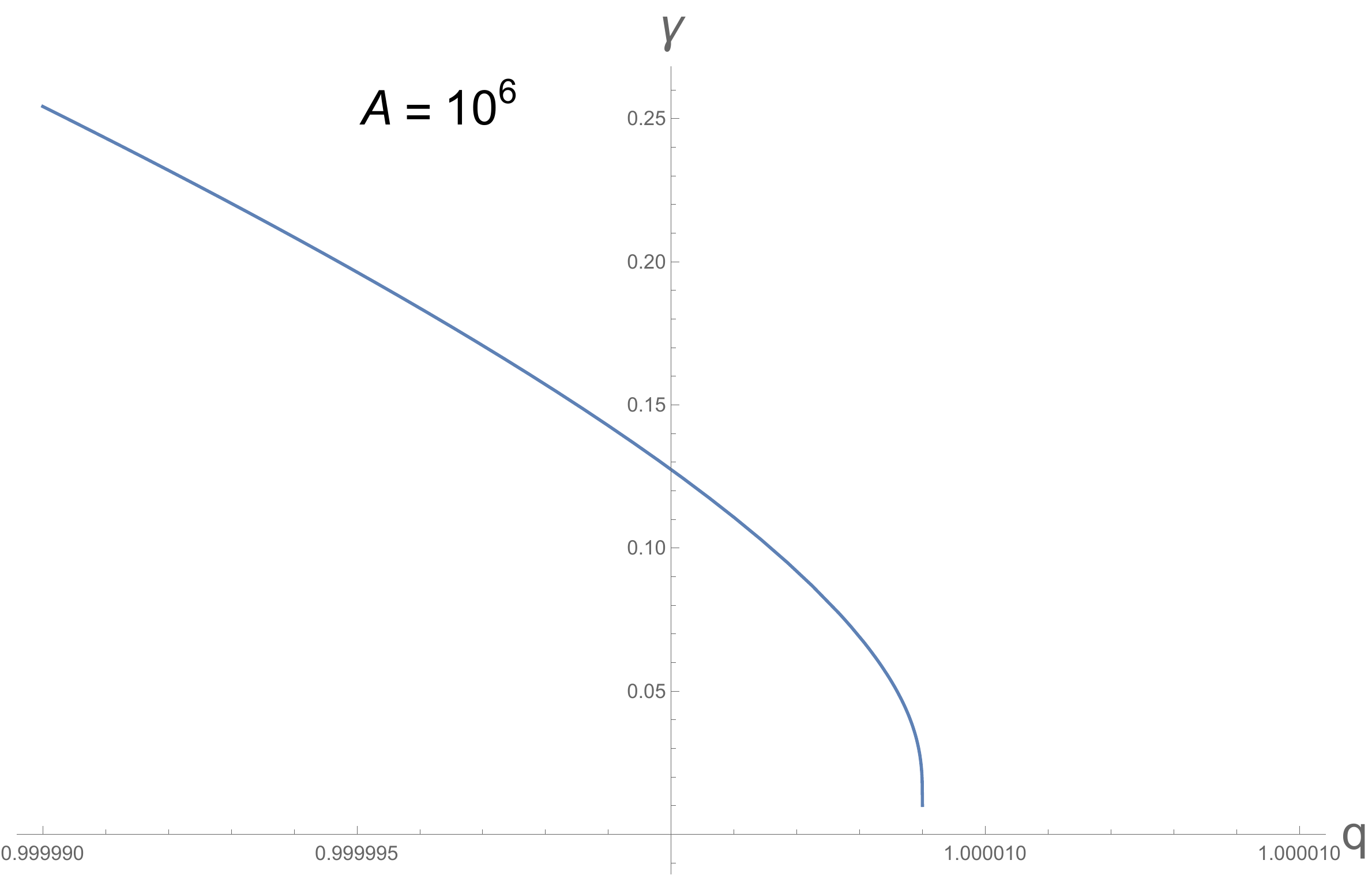}
\caption{\label{gbou}In this plot we have shown a highly resolved version of fig.(\ref{A106}) to zoom into the region $q>1$ to reveal its nature as $\g\to0$. It also shows clearly that, at $q=1$, $\g=\ln 2/\pi\sqrt 3=0.12738\cdots$.} 
\end{center}
\end{figure}

{\it Anomalous behaviour of $\g$:} However, there is a small section of the curve where $q>1$ and $\g<\ln 2/\pi\sqrt 3$ as can be seen from fig.(\ref{gbou}). In this regime, the above discussed physics fails because  $\g$ decreases with the departure of $q$ from unity. This implies that the coupling of the horizon with the bulk becomes weaker. But there is still a bias in the microstates due to $q\neq1$, although in the opposite sense i.e. now the more probable microstates appear more frequently. Hence, this section of the curve for $q>1$ represents some sort of anomaly which we are unable to address in the present discussion. We suspect that some explanation may come out if we study the underlying dynamics of the punctures in a similar line as in \cite{beck}. 
Further, an analysis of the eq.(\ref{gq}) shows that $\g\to0$ as $q\to(1+4/A)$. %i.e.
%\ba
%\lim_{q\to(1+4/A)}\gamma=\f{\ln 2}{\pi\sqrt 3}.\f{(-1)}{(-\infty)}=0.
%\ea
This shows that we can have values of $\g$ arbitrarily close to zero; but $\g$ can not be exactly equal to zero. This is physically consistent with the fact that the area spectrum vanishes for $\g=0$ which does not hold any meaning. Although this particular result may seem to be too trivial to need an explanation, but this analytic exploration is done just to ensure the behaviour of $\g$ in this limit as it is not evident from the plot in fig.(\ref{gbou})(all the plots have been drawn in Mathematica v10). 

Therefore, we can conclude that {\it we have obtained the Bekenstein-Hawking area law for black holes in LQG for arbitrary real positive values of $\g$}. %On the other hand, although in a restricted regime of validity($q\leq1$), {\it we have given a physical interpretation for the parameter $q$ of non-extensive statistical mechanics in the context of quantum gravity.}

\section{When $\g$ is experimentally determined}
Unlike the classical dynamics, the quantum dynamics is affected by $\g$, as it enters the quantum observables such as the area operator. Hence, the BHAL should follow from the microscopic degrees of freedom for all values of $\g$. In this context, it was argued in \cite{jac} that there could be a running of $\g$, the Newton's constant and an effective renormalization of the area in an effective field theory limit that may result in the BHAL for all values of $\g$. In this picture it might be possible that $\g$ has a non-trivial area dependence. However, whether this flow of $\g$ is connected or not with the one discussed here, can only be seen from the dynamics of the theory that is unavailable currently. In eq.(\ref{gq}), $\g$ has been expressed as a function of $q$ and $A$ so that one can figure out the range of its allowance from the plots i.e. $0<\g<\infty$ and also to explain the relation between the coupling strength of the bulk with the horizon and the bias in the microstates, when $\g$ is {\it unknown}. 
In principle, we should  not be able to determine $\g$ theoretically, like the $\theta$-parameter of quantum chromodynamics\cite{qcd1}. To be more elaborate, as mentioned earlier, $\g$ does not affect the classical dynamics of the theory. The classical equations of motion of general relativity are valid for any $\g$. Hence, the semiclassical BHAL should be valid irrespective of the value of $\g$. Only the fact that $\g$ appears as a multiplicative factor in the area spectrum, restricts it to take real and positive values. The graphical study of eq.(\ref{gq}) shows that the present exercise fulfill all these criteria.

Now, once we can determine $\g$ experimentally to be $\g_{exp}$(say), then eq.(\ref{gq}) will only give a relation between $q$ and $A$ given by   
\ba
\ln[1+\frac{A}{4}(1-q)]\cdot\g_{exp}=\frac{\ln 2 }{\pi\sqrt 3}\cdot\frac{A}{4}(1-q)\label{gq1}
\ea
Hence, given the area$(A)$ of a black hole(can be computed from the corresponding metric) one can find the value of the $q$ parameter from the solution of the above transcendental equation. Hence, the underlying statistics giving rise to the BHAL will be completely determined by the area of the black hole. Of course, the statistics will not be the Maxwell-Boltzmann statistics, but will be the Tsallis statistics whose $q$-parameter is determined by $A$ through eq.(\ref{gq1}). But, to see that explicitly, one needs to consider all possible spins and find out the distribution using the $q$-entropy as the starting point.

The fact that the parameter $q$ has now become a function of $A$ is physically well justified. It can be explained as follows. The $q$ parameter in the definition of the entropy takes into account the effect of the coupling between the horizon and the bulk as a bias in the microstates, as we have already discussed. Now, if we look at the coupling constant or the level of the CS theory, it is now given by $k=A/4\pi\g_{exp}$. So, the coupling strength between the horizon and the bulk is determined by $A$. Hence, it is expected that the bias in the microstates, represented by $q$, will also depend on $A$. This justifies the $A$ dependence of $q$ given by eq.(\ref{gq1}).

\section{ Conclusion} 
In course of obtaining the BHAL for black holes in LQG for arbitrary real positive $\g$, here we have set all the spins to $1/2$ for the horizon punctures as a first step of application of NESM in this context. Hence, one can doubt that the inclusion of all possible spins may spoil the nice interplay between $\g$ and $q$. In order to address this issue, it may be pointed out that $q=1$ brings us back to the scenario of unbiased microstates. In this case, the inclusion of all possible spins only change the value of $\g$ that yields the BHAL(see \cite{sigma} and the references therein). That is, one should expect that the graph in fig.(\ref{gbou}) cuts the ordinate at a slightly different point keeping the overall behaviour same. Nevertheless, to verify this fact quantitatively may prove to be a mathematically non-trivial problem, which we have left out for future study. 

Further, it may be mentioned that in \cite{jac} it was pointed out that the BHAL should be obtained for arbitrary $\g$, which means $\g$ could be negative and also complex. In this context, it may be pointed out that, since $\g$ appears as a multiplicative constant in the area spectrum, $\g<0$ gives negative quantum area of a black hole. This is completely unphysical. Hence, the very fact that we do not have $\g<0$ as a solution, only strengthens our case of application of NESM to obtain black hole entropy from LQG microstates. This is certainly an advantage over previous attempts to obtain black hole entropy from LQG for arbitrary $\g$\cite{gtune}. However, we are unable to address the case of $\g=\pm i$. This is worth mentioning because only $\g=\pm i$ gives a geometrical meaning to the connection variables in the four dimensional Lorentzian spacetime\cite{sam}. But, we fail to construct a well defined Hilbert space and also the area spectrum becomes imaginary, that we do not know how to explain.
It will be interesting to shed some light on the problem of black hole entropy calculation from LQG using NESM for $\g=\pm i$ if we can ever solve the above issues. Apart from this, from a broader perspective, this work establishes a key link between the microscopic theory of black holes in LQG and the phenomenology associated with the application of NESM to black hole thermodynamics problems\cite{1,2,3}. We hope that our results will lead to further innovative works in this direction.

Finally, let us end with the following, perhaps the most crucial, remark. The derivation of the BHAL from LQG microstates for arbitrary real positive $\g$ has been possible at the expense of the introduction of a new parameter in the definition of entropy. At the end, this parameter has to depend on the area of the black hole in a specific way so that we obtain the BHAL . Thus, the definition of the entropy depends on the area of the black hole and hence, in this sense, the definition is model dependent\cite{daniel}. Thus, in reality, the problem has been shifted from LQG to the definition of entropy\cite{hernando}. In our view point, it is better to keep the problem to the side which we do not understand very well i.e. definition of entropy, especially when it is being applied to a quantum gravity scenario, than to use a standard definition of entropy, considering it as sacrosanct and keep on the stance that there is a problem with a well-defined quantum theory as LQG. Rest is up to the reader to decide.

{\bf Acknowledgments:} I am grateful to A. Bravetti, L. F. Escamilla-Herrera and F. Nettel for giving me useful suggestions.  I want to thank V. G. Czinner for pointing out the references \cite{1,2,3}.  Also, I want to thank an anonymous referee for giving some useful suggestions. This work was funded by DGAPA fellowship of UNAM.


\begin{thebibliography}{77}
\bibitem{gtune} There were two attempts in the past in \cite{bia} and in \cite{gp} to obtain black hole entropy in LQG for arbitrary $\g$. However, both the derivations were valid only for local observers infinitesimally close and stationary with respect to the horizon. If we consider the case of \cite{bia}, there is no reason to assume that the local observer should  measure the entropy to be $A/4$ because the factor $1/4$ in the BHAL was fixed by Hawking for asymptotic observers\cite{haw}. Added to this, in \cite{bia}, $\g<0$ was allowed which contradicts the fact that the area of the horizon can not be negative. None of these problems arise in the present discussion. On the other hand, it was shown in \cite{am} that the claim made in \cite{gp} about obtaining black hole entropy for arbitrary $\g$ was not quite correct; $\g$ had a positive lower bound.
\end{thebibliography}
\end{document}